\documentclass[sigconf,natbib=true]{acmart}
\usepackage{tabularx}

\definecolor{gray}{HTML}{808080}
\definecolor{teal}{HTML}{21908C}
\definecolor{yellow}{HTML}{FDE725}
\definecolor{blue}{HTML}{1E2BC5}
\definecolor{purple}{HTML}{440154}
\definecolor{green}{HTML}{238E12}

\newtoggle{comments}
\toggletrue{comments}
\iftoggle{comments}{
  \newcommand{\notejz}[1]
    {{\color{teal}[{\bf JZ:} #1]}}
  \newcommand{\notewz}[1]
    {{\color{orange}[{\bf Weijie:} #1]}}
  \newcommand{\notetj}[1]
    {{\color{blue}[{\bf Thorsten:} #1]}}
  \newcommand{\notedt}[1]
    {{\color{green}[{\bf Doug:} #1]}}
  \newcommand{\todo}[1]
    {{\color{red}[{\bf TODO:} #1]}} 
  \newcommand{\cut}[1]
    {{\color{red}\sout{#1}}}

}{
  \newcommand{\notejz}[1]{}
  \newcommand{\notewz}[1]{}
  \newcommand{\notetj}[1]{}
  \newcommand{\notedt}[1]{}
  \newcommand{\todo}[1]{}
  \newcommand{\cut}[1]{}

}

\newcommand{\sysname}{\mbox{\sc {SteerEval}}}

\AtBeginDocument{%
  }

\setcopyright{acmlicensed}
\copyrightyear{2018}
\acmYear{2018}
\acmDOI{XXXXXXX.XXXXXXX}
\acmConference[Conference acronym 'XX]{Make sure to enter the correct
  conference title from your rights confirmation email}{June 03--05,
  2018}{Woodstock, NY}
\acmISBN{978-1-4503-XXXX-X/2018/06}

\begin{document}

%\title{ReccRevised: Measuring the Steerability of Natural Language Recommendations on User-Centered Tasks}
%%\title{RecRevised: Measuring the Steerability of Natural-Language Profiles for Recommendation} 

\title{SteerEval: A Framework for Evaluating Steerability with Natural Language Profiles for Recommendation}

% \affiliation{%

\author{Joyce Zhou}
\affiliation{%
  \institution{Cornell University}
  \city{Ithaca}
  \state{New York}
  \country{USA}
}
\email{jz549@cornell.edu}
\orcid{0000-0003-1205-3970}

\author{Weijie Zhou}
\affiliation{%
  \institution{Cornell University}
  \city{Ithaca}
  \state{New York}
  \country{USA}
}
\orcid{0009-0004-7721-0150}

\author{Doug Turnbull}
\affiliation{%
  \institution{Ithaca College}
  \city{Ithaca}
  \state{New York}
  \country{USA}
}
\orcid{0009-0001-7252-1855}

\author{Thorsten Joachims}
\affiliation{%
  \institution{Cornell University}
  \city{Ithaca}
  \state{New York}
  \country{USA}
}
\orcid{0000-0003-3654-3683}

\renewcommand{\shortauthors}{Zhou et al.}

\begin{abstract}
%This is a weak opening sentence, so let me rephrase: In recent years, there has been growing interest in centering recommendation around natural language profiles to represent user preferences, with work demonstrating how accurate, interpretable, and steerable such profiles may be.
%(e.g., aspirational, specific, context-dependent)

%Users profiles represented as natural-language text have recently attracted attention not only for their improved interpretability, but also for their potential to make recommender systems more steerable. Natural‑language profiles allow users to directly edit their profiles to explicitly articulate preferences that maybe difficult to infer from past behavior. 

Natural-language user profiles have recently attracted attention not only for improved interpretability, but also for their potential to make recommender systems more steerable. By enabling direct editing, natural-language profiles allow users to explicitly articulate preferences that may be difficult to infer from past behavior.
However, it remains unclear whether current natural‐language-based recommendation methods can follow such steering commands. 
While existing steerability evaluations have shown some success for well-recognized item attributes (e.g., movie genres), we argue that these benchmarks fail to capture the richer forms of user control that motivate steerable recommendations.
%However, existing steerability evaluations largely focus on well-recognized item attributes.
%We argue that this fails to capture the nuance of potential user control that motivates steerable natural language profile recommendations.
To address this gap, we introduce \sysname, an evaluation framework designed to measure more nuanced and diverse forms of steerability by using interventions that range from genres to content-warning for movies.
We assess the steerability of a family of pretrained natural-language recommenders, examine the potential and limitations of steering on relatively niche topics, and compare how different profile and recommendation interventions impact steering effectiveness.
% We find that while recommenders based on embedding models perform poorly on some steering tasks, most recommenders are fairly resilient across tasks and steered topics, although they often lack appropriate world knowledge.
Finally, we offer practical design suggestions informed by our findings and discuss future steps in steerable recommender design.
\end{abstract}



\begin{teaserfigure}
  \begin{center}
    \includegraphics[width=0.8\textwidth]{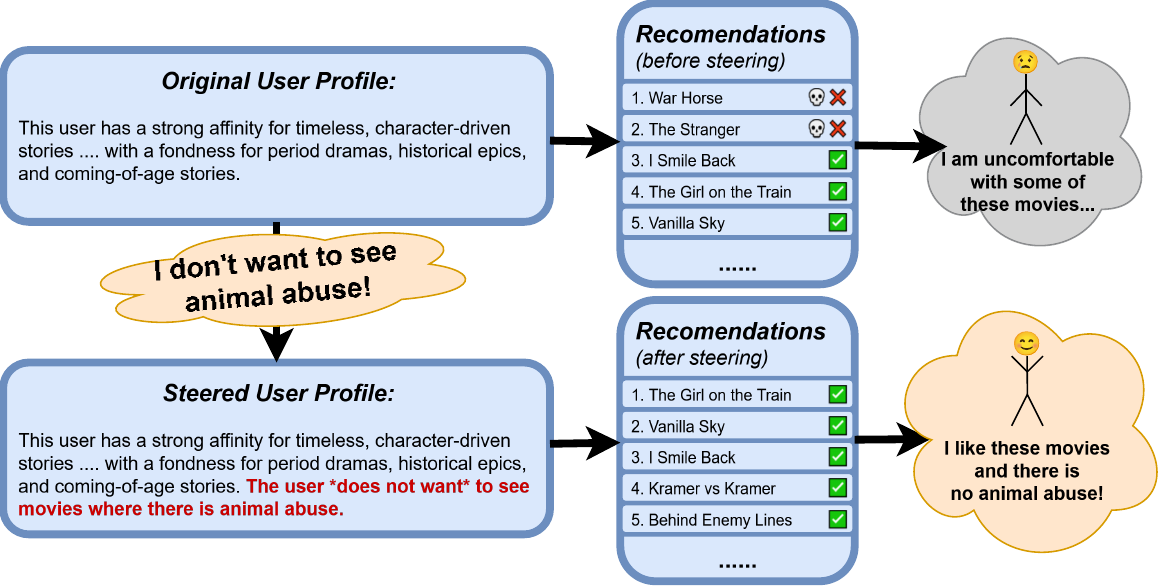}
  \end{center}
  \caption{
  A motivating use case for steerable recommendation, illustrating how a user can explicitly articulate interests or sensitivities (e.g., avoiding a trigger) using natural language to \emph{steer} the system toward better-aligned recommendations.
  %An illustration of the steering motivations that the \sysname\ evaluation framework captures.
  % An illustration of the \sysname\ steerability evaluation pipeline.
  % For a given original user profile and a desired steering action, the steering intervention produces an updated steered profile, and a ranking method produces recommendations for each profile version.
  }
  \Description{
  A figure that shows an example pipeline where the user originally has some natural language preference text (Original User Profile) and exclaims "I don't want to see animal abuse!". Based on this, the original profile is revised to produce an updated, Steered User Profile. Both the Original and Steered User Profiles are each used to produce a recommendation list. The Original User Profile produces an example ranking with 2 out of the top 5 movies containing animal abuse, and the user is sad when seeing this ranking. The Steered User Profile produces an example ranking with 0 out of the top 5 movies containing animal abuse, and the user is happier when seeing this ranking, saying that "I like these movies and there is no animal abuse".
  % A figure that shows an example pipeline where the user originally has some natural language preference text (Original User Profile) and exclaims "I don't want to see animal abuse!". Based on this, the original profile is passed through a Steering Intervention component with several possible intervention methods to produce an updated, Steered User Profile. Both the Original and Steered User Profiles are then passed through the same type of ranking component, and each produces a recommendation list. The Original User Profile produces an example ranking with 2 out of the top 5 movies containing animal abuse. The Steered User Profile produces an example ranking with 0 out of the top 5 movies containing animal abuse.
  }
  \label{fig:teaser}
\end{teaserfigure}

\received{20 February 2007}
\received[revised]{12 March 2009}
\received[accepted]{5 June 2009}

\maketitle

\section{Introduction}
\label{sec:introduction}

%With the increasing prevalence of large language models in user interface and system design, there has also been increasing interest in using natural text to encode user preferences for recommendation systems. Work in this area has shown that recommendation systems centered around natural language profiles demonstrate competent sequential recommendation performance, while enabling affordances such as end-user transparency, interpretability, scrutability, and recommendation controllability.

The availability of large language models has created new opportunities for using natural-language text to articulate user preferences for recommendation. Recommender systems that leverage natural-language profiles have demonstrated competent recommendation performance \cite{sannerLargeLanguageModels2023, gaoLangPTuneOptimizingLanguagebased2025}, while also enabling promising new affordances such as end-user transparency, scrutability, and \emph{steerability} \cite{radlinskiNaturalLanguageUser2022, penalozaTEARSTextualRepresentations2024, ramosTransparentScrutableRecommendations2024}.

%As natural language profiles encode user preferences into some human-readable text that is directly used by large language models for downstream recommendations, it stands to reason that users should be able to read, understand, and self-reflect on these texts. In brief, this is \textit{scrutability}: the ability for an end user to understand their profile. A scrutable profile opens up the possibility that users can edit a profile for any number of reasons, including but not limited to changing their recommendation habits, setting requests that are difficult to otherwise articulate, or adjusting to a different audience than usual. This is \textit{steerability}: the ability of the recommendation system to adequately update recommendations based on the end user's edits. We are particularly interested in steerability (which is also called controllability in other works).

As natural-language profiles encode user preferences into human-readable text that is directly consumed by large language models for downstream recommendations, they allow users to read, interpret, and reflect on these texts.
Furthermore, natural-language profiles enable users to edit their preferences for a variety of reasons, including changing their recommendation habits (e.g., more aspirational), expressing preferences that are otherwise difficult to articulate (e.g,. specific sensitivities as shown in Figure \ref{fig:teaser}), or tailoring their profile to a different context (e.g., watching a movie with children).
This affords a new form of {\em steerability}: the ability of the recommender system to appropriately update its recommendations in response to user edits to their natural-language profile. 

%With the increased flexibility of natural language profiles to represent user interests and disinterests, also comes increased diversity in the ways that users may want to edit them. While past works have included evaluations of recommendation steerability through profile updates, these largely focused on a limited set of well-known attributes, such as movie genre. While this is a valuable first step, it fails to capture how large the space of potential user steering actions or methods is, and how natural language profiles may enable these broader steering actions. Ideally, we want to allow users to steer the recommendation system towards or away from any set of items, attributes, diversity, or other preferences as they so choose, even if those preferences are relatively uncommon.

Furthermore, as natural‑language profiles provide greater flexibility in expressing user interests and disinterests, the range of ways in which users may wish to edit them expands as well. Although prior work has evaluated recommendation steerability through profile updates, these evaluations have typically centered on a narrow set of well‑known attributes, such as movie genres. While this is a valuable starting point, it overlooks the vast space of potential steering actions and methods that natural‑language profiles can support. In principle, users should be able to steer a recommendation system toward or away from any collection of items, attributes, diversity goals, or other preferences they choose, even if those preferences are relatively specific or uncommon.

In this paper, we present \sysname, a comprehensive steerability evaluation framework and benchmark designed for natural-language profile recommendation systems. We focus on the movie domain, but the framework is extendable to more steering actions, datasets, and domains such as music \cite{gaglianoUsingLanguageModels2025} and scientific papers \cite{arustashviliSciNUPNaturalLanguage2025}.
We show an illustration of the evaluation pipeline in Figure \ref{fig:pipeline_detail}.

A key resource provided by \sysname\ is a benchmark dataset of testable steering topics around multiple sources of human-readable, important, and relevant potential goals. We demonstrate how this benchmark can be used to compare potential recommender methods, steering directions, steerable topics, and steering methods.
We find that while some attributes such as ranking methodology heavily impact steerability, in general natural-language-profile recommenders are successfully steerable and quite resilient to different steering action variations.
However, we find that language models often lack relevant world knowledge, which makes some topics more difficult to steer.
Finally, we present practical design suggestions based on our results and discuss future steps.

Our implementation of the \sysname\ framework and the dataset we created is available online\footnote{\url{https://github.com/cephcyn/SteerEval}}.

% Our contributions are:
% \begin{itemize}
%     \item We present \sysname, an offline evaluation suite for natural language recommendation steerability, with a modular design that can be extended beyond the movie recommendation domain and potentially beyond natural language profile recommenders.
%     \item We build a steerability evaluation dataset based on multiple existing movie datasets.
%     \item We evaluate a family of pretrained natural language profile recommendation systems using \sysname, using a variety of steering actions, topics, and intervention methods. Based on this, we discuss design suggestions, current shortcomings, and potential improvements for both steerable recommendation and evaluation suite design.
% \end{itemize}

\section{Background}
\label{sec:background}

\subsection{Natural language profile recommendation}
\label{sec:background_profile_recc}

A natural-language-profile recommender is a recommender system that uses a natural language text that describes a user's preferences in order to produce recommendations \cite{radlinskiNaturalLanguageUser2022, balogTransparentScrutableExplainable2019}.

In contrast to traditional uninterpretable vector representations of users or items \cite{radlinskiNaturalLanguageUser2022}, a major motivation for natural language profile recommenders is their scrutability, transparency, explainability, and potential steerability.
A user profile text can be created through any number of ways, although most existing works have generated profile texts by extracting keywords from reviews \cite{torbatiRecommendationsConciseUser2025} or prompting a large language model with \cite{zhouLanguageBasedUserProfiles2024, gaoLangPTuneOptimizingLanguagebased2025, penalozaTEARSTextualRepresentations2024, ramosTransparentScrutableRecommendations2024, radlinskiNaturalLanguageUser2022} metadata based on user interaction history.
Similarly, the user profile text can be used to generate recommendations in a variety of ways, including large language model prompting \cite{radlinskiNaturalLanguageUser2022, ramosTransparentScrutableRecommendations2024} or embedding-based decoding \cite{torbatiRecommendationsConciseUser2025, zhouLanguageBasedUserProfiles2024, penalozaTEARSTextualRepresentations2024, gaoLangPTuneOptimizingLanguagebased2025}.

Most existing works on natural language profile recommenders have found that they show recommendation accuracy on par with or sometimes better than existing systems \cite{gaoLangPTuneOptimizingLanguagebased2025, penalozaTEARSTextualRepresentations2024}, may be especially advantageous in cold-start scenarios \cite{sannerLargeLanguageModels2023, zhouLanguageBasedUserProfiles2024, ramosTransparentScrutableRecommendations2024} and can be improved with appropriate profile generation \cite{gaoLangPTuneOptimizingLanguagebased2025} or profile decoder \cite{ramosTransparentScrutableRecommendations2024, penalozaTEARSTextualRepresentations2024, gaoLangPTuneOptimizingLanguagebased2025} model finetuning.
However, the main intended benefit is still end-user transparency and scrutability.

\subsection{Recommendation steerability}
\label{sec:background_steerability}

Steerability (also called controllability \cite{wangUsercontrollableRecommendationFilter2022, parraUsercontrollablePersonalizationCase2015, mysoreEditableUserProfiles2023} or configurability \cite{harperPuttingUsersControl2015}) is the ability of an end user to control the behavior of a system \cite{deanRecommendationsUserAgency2020, greenGeneratingTransparentSteerable2009}.
Steerable recommender systems enable end users to modify what recommendations they receive \cite{harperPuttingUsersControl2015, baoDiff4SteerSteerableDiffusion2025}.
Steering can take many forms, most of which depend on the recommender system architecture itself.
For example, a recommender could be considered steerable if it allows users to select or adjust weighting for item keywords or tags \cite{mysoreEditableUserProfiles2023, greenGeneratingTransparentSteerable2009}, adjust recommendation hyperparameters \cite{harperPuttingUsersControl2015, parraUsercontrollablePersonalizationCase2015}, or allows item history revisions to reflect desired changes as appropriate \cite{deanRecommendationsUserAgency2020}.
Previous work in natural language profile recommendation has largely assumed that a user can freely edit the natural language profile to update downstream recommendations \cite{radlinskiNaturalLanguageUser2022}.

Ideally, a steerable recommender system is scrutable so the user understands how best to make changes, allows broad support for the changes that users do want, and does not require too much effort to steer \cite{deanRecommendationsUserAgency2020}.
While many existing recommender systems designed to support steerability recognize sets of steering actions based on formalized metadata \cite{wangUsercontrollableRecommendationFilter2022}, large language models ideally enable natural-language-profile recommenders to respond to more nuanced user instructions and commands as well \cite{radlinskiNaturalLanguageUser2022}.
User motivations for steering a recommender might include fixing incorrect information \cite{tintarevSurveyExplanationsRecommender2007}, introducing new preferences \cite{parraUsercontrollablePersonalizationCase2015, radlinskiNaturalLanguageUser2022}, giving context-specific instructions \cite{radlinskiNaturalLanguageUser2022}, diversifying from filter bubbles \cite{kaliraiYouTodayBetter2024, deanRecommendationsUserAgency2020, wangUsercontrollableRecommendationFilter2022}, or setting aspirational or long-term goals for self-actualization \cite{kaliraiYouTodayBetter2024, yangHowIntentionInformed2019, knijnenburgRecommenderSystemsSelfActualization2016, liangEnablingGoalFocusedExploration2023}.

\subsection{Evaluating steerability}
\label{sec:background_steerability_evaluation}

There are a number of previous works that propose and evaluate the usability of recommendation steering interfaces, including keyword or tag-based interfaces \cite{yangHowIntentionInformed2019, mysoreEditableUserProfiles2023, greenGeneratingTransparentSteerable2009} and conversational recommendation \cite{harperPuttingUsersControl2015}.
These evaluations often prioritize user satisfaction \cite{harperPuttingUsersControl2015, yangHowIntentionInformed2019, mysoreEditableUserProfiles2023} or ease of use \cite{parraUsercontrollablePersonalizationCase2015}.

\begin{figure*}[h]
  \centering
  \includegraphics[width=\linewidth]{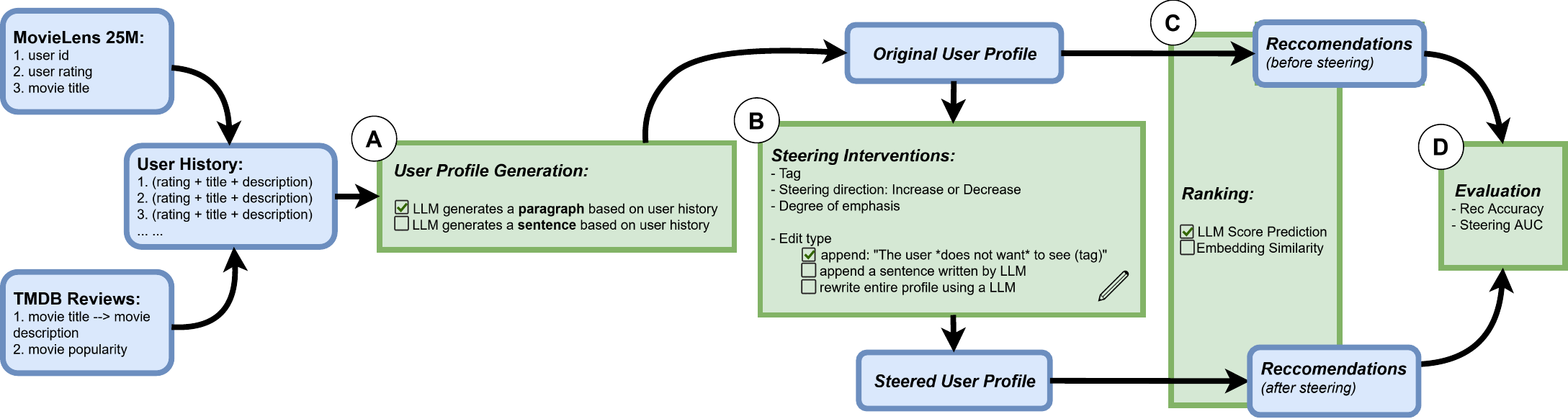}
  \caption{
  An illustration of the \sysname\ steerability evaluation pipeline interwoven with the natural language recommendation process.
  The user profile generation process (A) creates an original user profile based on item metadata and user rating history (see section \ref{sec:design_recommendation_creation}). For a given original user profile and a desired steering action, the steering intervention (B) produces an updated steered profile (see section \ref{sec:design_steering_intervention}), and a ranking method (C) produces recommendations for each profile version (see section \ref{sec:design_recommendation_ranking}). Each pair of rankings is evaluated (D) for accuracy and steering effectiveness (see section \ref{sec:design_steering_evaluation}).
  Checkmarks denote the methods that we select for this example using our ``default'' pipeline.
  }
  \Description{
  A figure illustrating the STEEREVAL evaluation pipeline for steerable recommendation. MovieLens user ratings and IMDb metadata are combined to form a user history, which is used in step (A) to generate an Original User Profile using a language model. In step (B), a steering intervention modifies the profile based on a desired steering action, producing a Steered User Profile. In step (C), both the Original and Steered User Profiles are passed to the same ranking methods to generate recommendations before and after steering. In step (D), the two recommendation lists are evaluated for recommendation accuracy and steering effectiveness. Checkmarks indicate the specific methods selected in the default pipeline.
  }
  \label{fig:pipeline_detail}
\end{figure*}

Previous works have also done some interpretability \cite{gaoLangPTuneOptimizingLanguagebased2025, radlinskiNaturalLanguageUser2022} and steerability \cite{penalozaTEARSTextualRepresentations2024, ramosTransparentScrutableRecommendations2024} evaluations to demonstrate how natural language profile recommenders are responsive to naturally phrased changes to end user preferences.
These measure steerability by identifying a target action and relevant item(s) that should be affected by the action, editing the profile in a way that reflects this action, then measuring changes between the original and updated item rankings for the relevant item(s).

Existing natural language profile recommendation steerability evaluations have largely focused on metadata for a very limited and well-known set of potential steering actions, such as genre \cite{penalozaTEARSTextualRepresentations2024}.
While this is an understandable limitation, it does not demonstrate the advantage that a natural language profile recommender may have over an existing system that explicitly supports steering (e.g., filtering, post-hoc score adjustment) on specific metadata tags (e.g., genres, movie director, release year).
While natural language profile recommendations might be able to support more nuanced steering actions (more nuanced goals and/or more niche subjects), evaluating on commonly known genre tags fails to adequately capture this strength.
Other steerability evaluations have attempted to expand the set of steering actions by extracting notable keywords from user-written reviews to get a broader range of the features that users may care about or specify when revising a profile \cite{ramosTransparentScrutableRecommendations2024}.
% \todo{expand to steering action intent and method?}
We further expand on this by examining a wider range of user-specified important annotations and explore how users might use different strategies when attempting to convey the steering action.

\section{\sysname\ Pipeline Design}
\label{sec:design}

We now explain the design of \sysname\ and the variations in natural language profile recommenders that we test.
Figure \ref{fig:pipeline_detail} depicts the full natural language profile recommender pipeline interwoven with \sysname\ evaluation components.
In this figure, we identify four major components: profile creation, steering intervention, item ranking, and evaluation.

% The natural language recommender components (sections A and C in Figure \ref{fig:pipeline_detail}: user profile generation and ranking) are discussed in section \ref{sec:design_recommendation}.
% The steering evaluation components (sections B and D in Figure \ref{fig:pipeline_detail}: steering intervention and evaluation) are discussed in section \ref{sec:design_steering}.

\subsection{Recommendation}
\label{sec:design_recommendation}

There is a variety of proposed natural language profile recommender methods, each consisting of two major components: natural language profile creation and item ranking.
To capture the variations in recommendation pipeline design, we test a range of pretrained profile creation and item ranking methods inspired by the core design ideas from existing work.

\subsubsection{Profile creation}
\label{sec:design_recommendation_creation}

In order to provide recommendations based on a natural language profile, the system must first create a profile.
This is represented by section A in Figure \ref{fig:pipeline_detail}.

Previously proposed profile creation methods largely prompt a LLM with information based on user watch history \cite{zhouLanguageBasedUserProfiles2024, gaoLangPTuneOptimizingLanguagebased2025, penalozaTEARSTextualRepresentations2024} or aggregated preference keywords \cite{ramosTransparentScrutableRecommendations2024}, potentially with some type of model finetuning involved \cite{gaoLangPTuneOptimizingLanguagebased2025}.
These are all intended to produce preference summaries in natural language prose. For our experiments, we do not consider profiles that consist of keyword or tag collections \cite{mysoreEditableUserProfiles2023}.

For our experiments, we consider two profile creation methods that differ in the length of the text:
\begin{itemize}
    \item Prompt a LLM to generate a short paragraph
    \item Prompt a LLM to generate a single sentence
\end{itemize}
which are both based on the same watch history and item metadata.

\subsubsection{Item ranking}
\label{sec:design_recommendation_ranking}

Finally, there have been a variety of methods suggested for how to generate recommendations using a natural language profile.
Given some original or steered profile and a collection of items to rank, the ranking method collects information about each item and ideally produces an accurate or appropriately updated ranking for the user.
This is represented by section C in Figure \ref{fig:pipeline_detail}.

In our experiments, we consider two ranking methods:
\begin{itemize}
    \item Calculate embedding similarity between the natural language profile and item metadata texts \cite{gaoLangPTuneOptimizingLanguagebased2025, penalozaTEARSTextualRepresentations2024}
    \item Prompt a LLM to predict the score for each item based on the profile and item metadata texts \cite{ramosTransparentScrutableRecommendations2024, zhouLanguageBasedUserProfiles2024}
    % \item Prompt a LLM to produce an ordered list given the profile and a collection of all items to rank \jznote{we do not do this because of context length issues}
\end{itemize}
The metadata text for each item contains movie title and TMDb description.

\subsection{Steering Intervention}
\label{sec:design_steering}

We consider a system steerable if it responds appropriately to a user's instructions.
A successful steering action is one where a user can describe their updated preferences, and items that are relevant to the change will be moved in the recommendations (e.g., ranking) as desired.
We test whether a system can respond to a diverse range of steering intentions by using varying topics, steering tasks, and steering methods to emulate diversity in user interests and how they might edit their natural language profile.

% \subsubsection{Steering topics}
\label{sec:design_steering_topic}

Unsurprisingly, there are many possible preferences and ways to express them. Moreover, many concepts have ambiguous relevance criteria, making steering success difficult to evaluate.
To overcome this, we use existing movie annotation databases to select steerable tags and determine whether a movie is relevant to the tag or not.
We use these tag datasets:
\begin{itemize}
    \item \textbf{Genre}: The MovieLens-25M \cite{harperMovieLensDatasetsHistory2015} dataset contains a collection of 19 total genre tags. Most movies are linked to one or more genre. We consider a movie "relevant" if it is linked to a genre, and "irrelevant" otherwise.
    \item \textbf{Trigger}: \emph{Does The Dog Die}\footnote{https://www.doesthedogdie.com/} is a website that collects crowd-sourced answers to a curated list of sensitive topic and trigger warning questions. We use the NavigatingSensitivity 2025 dataset \cite{kovacsDatasetsNavigatingSensitive2025}, which collects aggregated information for 137 total trigger warnings.
    128 of these tags are common enough to be used for our experiments.
    We consider a movie "relevant" if over 75\% of votes say that a trigger warning applies. We algorithmically rename these tags for grammatical convenience in steering.
    We observed that some of these trigger warnings are often misinterpreted or rejected by pretrained large language models, likely due to model safety guardrails. We exclude these rejected tags from our experiments and discuss this more in Section \ref{sec:experiment_per_topic}.
    After this filtering, we test a total of 75 trigger warning tags.
    % \item MovieLens tag genome, which are user-submitted free text entries that get scored on how relevant they are to each movie
    % \item TMDb user-submitted tags. These are not currently used.
\end{itemize}

% \subsubsection{Steering types}
\label{sec:design_steering_type}

% We also test two simple steering task types based on an informal sample:
We also test two steering direction tasks:
\begin{itemize}
    \item A user may want to \textbf{increase} how often they see some type of content. For instance, the user might be newly curious about a certain genre or topic.
    \item A user may want to \textbf{decrease} how often they see some type of content. For instance, the user might avoid certain sensitive topics.
    % \item increase/decrease with multiple tags?
    % \item "diversify"???
\end{itemize}

% \subsubsection{Steering intervention}
\label{sec:design_steering_intervention}

Given some intention to steer recommendations, users might edit a profile in any number of ways.
This edit should be human-readable, and ideally should not take too much effort to achieve.
As a whole, this is represented by section B in Figure \ref{fig:pipeline_detail}.
In our experiments, we consider these steering interventions, each of which take steered topic and intended action into account:
\begin{itemize}
    \item \textbf{Append a template sentence} to the original profile based on the steering action and tag name.
    \item \textbf{Append a LLM-generated sentence} based on the steering action and tag name.
    \item \textbf{Prompt a LLM to revise the original profile} based on the desired steering action \cite{ramosTransparentScrutableRecommendations2024, penalozaTEARSTextualRepresentations2024}
    % \notejz{While this was on average the most effective steering intervention (based on results discussed in Section \ref{sec:experiment_steering_method}), we argue that it is not the most realistic default action because it would take a larger amount of effort for an end user to replicate.}
    % \item Fully replace the profile \cite{penalozaTEARSTextualRepresentations2024}
\end{itemize}

While we work in the movie recommendation domain, the steering tasks and intervention methods can be adapted for other domains, with appropriate substitutions for tag datasets.

\subsubsection{Evaluation}
\label{sec:design_steering_evaluation}

To evaluate steerability, we compare an item ranking produced by the original unmodified profile against an item ranking produced by the steered profile.
This is represented by section D in Figure \ref{fig:pipeline_detail}.

To measure steering impact, we define a tag-based AUC metric $AUC_{t}$, which captures how highly rated items that are relevant to a tag $t$ tend to be within a ranking.
A ranking with all relevant items preceding all irrelevant items has ${AUC}_t = 1$, a ranking with all relevant items after all irrelevant items has ${AUC}_t = 0$, and a random ranking is expected to have ${AUC}_t = 0.5$.

We compute ${AUC}_t$ \cite{gaglianoUsingLanguageModels2025} and the change in ${AUC}_t$ ($\Delta AUC_{t}$) of targeted items \cite{penalozaTEARSTextualRepresentations2024, ramosTransparentScrutableRecommendations2024} between the original and post-steering rankings.
When doing an increase action, we want items that are relevant to the targeted tag to ascend the ranking, so higher (positive) $\Delta {AUC}_t$ is better.
When doing a decrease action, we want these items to descend, so lower (negative) $\Delta {AUC}_t$ is better.
% We also log accuracy, for the experiment in section \ref{sec:experiment_accuracy}.
% We also do replicate NDCG and recall metrics used in UCL and TEARS initial steerability evaluation experiments. See Appendix \ref{app:replication}.

% \notejz{define subgroup accuracy here?, see section \ref{sec:experiment_accuracy}}

% \begin{itemize}
%     \item Calculate delta of coverage@N for target set of items pre- and post-steering (UCL)
%     \item Calculate delta of NDCG@N for target set of items pre- and post-steering (TEARS)
%     \item Calculate delta of ranking of a single target item if present pre-and post-steering (TEARS)
%     \item Calculate AUCROC for target set of items pre- and post-steering (we primarily use this)
% \end{itemize}

\section{Experiments}
\label{sec:experiment}

We investigated how well different recommendation, profile creation, steering intervention, and item ranking methods support steering across all tags for both the increase and decrease tasks.
% \notejz{Probably worth noting that absolutely none of these pipeline components were finetuned for steerability OR recommendation performance itself - we are using pretrained models with a little bit of prompt engineering}

\subsection{Experiment setup}
\label{sec:experiment_setup}

In our experiments, we use the MovieLens 25M dataset for user rating history\footnote{\href{https://www.grouplens.org/datasets/movielens/25m/}{https://www.grouplens.org/datasets/movielens/25m/}} \cite{harperMovieLensDatasetsHistory2015}.
To produce rankings, we combine MovieLens movie title information with TMDb\footnote{\href{https://www.themoviedb.org/}{https://www.themoviedb.org/}} movie descriptions.
We filter out all movies which are missing a TMDb description.

We randomly sample users from this dataset with at least 100 total reviews.
We use the first 50 items from each user review history to generate the original user profile.
For each user, we perform a steering action by modifying their original profile to increase or decrease their preference for some tag.
For each user and steering action, we collect a set of 100 movies to rank, which contains exactly 50 items that are relevant to the steered tag and 50 non-relevant items.
To measure accuracy, exactly 1 of these 100 movies is the movie that the user watched next, which may be one of the relevant or non-relevant items.
To minimize popularity bias in ranking, the other 99 items are selected to minimize difference between their TMDb popularity score and that of the accurate next item.
We evaluate steering success by generating item rankings based on the original and steered profile for the same set of 100 items and measuring the change in how highly tag-relevant items are ranked.

We select steering intervention templates that are lightly prompt engineered to maximize steering effects when possible.
For the Genre tags, we use the template "Please show the user movies that satisfy: \textit{Mystery}." for increase and "The user *does not want* to see \textit{Mystery} movies." for decrease, where "\textit{Mystery}" is an example tag name.
For the Trigger tags, we use the template "The user *wants* to see movies where \textit{the dog dies}." for increase and "The user *does not want* to see movies where \textit{the dog dies}." for decrease, where "\textit{the dog dies}" is an example tag name.

We use pretrained Llama-3.1-8B-Instruct for all generative large language model components (e.g. profile creation, LLM-related steering intervention, and score prediction), and pretrained mxbai-embed-large-v1 for all embedding model components (e.g. embedding similarity ranking).
When using a LLM to predict scores, responses that are missing a score are treated as having predicted score 0.
Ties in predicted score rankings are randomly broken.

In total, for each pipeline variation and steering action (increase or decrease), we evaluate steering effectiveness across 94 tags\footnote{75 trigger tags and 19 genre tags} with 10 users each, for a total of 940 steering scenarios.
Result metrics are averaged uniformly across all 940 scenarios.
Computing 10 steering scenarios with the default pipeline takes approximately 40 minutes on one A6000 GPU.
Full profile generation prompts, steering intervention prompts, ranking prompts, templates, and example responses are available in Appendix \ref{app:prompts}.

\subsection{How do different recommender systems respond to steering?}
\label{sec:experiment_recommender_method}

We first explore how steerability varies across multiple configurations of natural language recommender components.
For simplicity and to emulate low-effort user actions, we perform a steering intervention by appending a template sentence based on steering task and tag name by default.

\begin{table}[]
    \centering
    \begin{tabular}{l|c c}
         & Incr. & Decr. \\
         Ranking method & $\Delta {AUC}_t$ ($\uparrow$) & $\Delta {AUC}_t$ ($\downarrow$) \\
         \midrule[0.15ex]
         % (Genre) & & \\
         % \textit{LLM scoring} & 0.1095 (*) & \textbf{-0.0654} (*) \\
         % Embedding & \textbf{0.1068} (*) & 0.0690 \\
         % \midrule[0.15ex]
         % (Trigger-all) & & \\
         % \textit{LLM scoring} & 0.0297 (*) & \textbf{-0.0169} (*) \\
         % Embedding & \textbf{0.0521} (*) & 0.0412 \\
         % \midrule[0.15ex]
         % (Trigger-allowed) & & \\
         % \textit{LLM scoring} & 0.0320 (*) & \textbf{-0.0168} (*) \\
         % Embedding & \textbf{0.0527} (*) & 0.0417 \\
         % \midrule[0.15ex]
         % (All) & & \\
         % \textit{LLM scoring} & 0.0400 (*) & \textbf{-0.0231} (*) \\
         % Embedding & \textbf{0.0592} (*) & 0.0448 \\
         % \midrule[0.15ex]
         % (All-allowed) & & \\
         LLM scoring & 0.0476 (*) & \textbf{-0.0266} (*) \\
         Embedding & \textbf{0.0636} (*) & 0.0473 \\
    \end{tabular}
    \caption{Ranking method comparison results for a paragraph-length profile and template appending intervention. The best-performing method for each task is bolded. Asterisks denote statistically significant steering success based on one-tail paired t-test ($\alpha=0.05$). Note that the embedding similarity ranker actually \textit{increases} tag presence when attempting the decrease task.}
    \label{tab:compare_ranker}
    \vspace{-2em}
\end{table}

\begin{table}[]
    \centering
    \begin{tabular}{l|c c}
         & Incr. & Decr. \\
         Profile generation & $\Delta {AUC}_t$ ($\uparrow$) & $\Delta {AUC}_t$ ($\downarrow$) \\
         \midrule[0.15ex]
         % (Genre) & & \\
         % \textit{Paragraph-length} & \textbf{0.1095} (*) & \textbf{-0.0654} (*) \\
         % Sentence-length & 0.1084 (*) & -0.0643 (*) \\
         % \midrule[0.15ex]
         % (Trigger-all) & & \\
         % \textit{Paragraph-length} & 0.0297 (*) & \textbf{-0.0169} (*) \\
         % Sentence-length & \textbf{0.0497} (*) & -0.0155 (*) \\
         % \midrule[0.15ex]
         % (Trigger) & & \\
         % \textit{Paragraph-length} & 0.0320 (*) & \textbf{-0.0168} (*) \\
         % Sentence-length & \textbf{0.0536} (*) & -0.0127 (*) \\
         % \midrule[0.15ex]
         % (All-all) & & \\
         % \textit{Paragraph-length} & \textbf{0.0400} (*) & \textbf{-0.0231} (*) \\
         % Sentence-length & 0.0573 (*) & -0.0218 (*) \\
         % \midrule[0.15ex]
         % (All) & & \\
         Paragraph-length & 0.0476 (*) & \textbf{-0.0266} (*) \\
         Sentence-length & \textbf{0.0647} (*) & -0.0231 (*) \\
    \end{tabular}
    \caption{Profile generation method comparison results for a LLM score ranker and template appending intervention. The best-performing method for each task is bolded. Asterisks denote statistically significant steering success based on one-tail paired t-test ($\alpha=0.05$). Sentence-length profiles perform well on the increase task, but are otherwise comparable to paragraph-length profiles.}
    \label{tab:compare_profiler}
    \vspace{-2em}
\end{table}

\label{sec:experiment_recommender_method_ranking}
We first examine the item ranking component.
Given some profile, a set of items to rank, and some item metadata, how does a recommender rank these items, and how does that ranking method influence how steerable the ranking is?

We compare the steering effectiveness between a LLM-prompted score prediction and a text embedding similarity ranker in Table \ref{tab:compare_ranker}.
We find that LLM scoring is broadly successful on both the increase and decrease tasks.
While embedding similarity was noticeably more effective on the increase task, it did consistently fail on the decrease task.
We suspect that this is because any explicit mention of a tag, whether liked or disliked, in the profile text would make its text embedding closer to items that are relevant to the tag.
That is, the embedding model seems to consider any positive or negative sentiment about a topic to be more similar than no mention at all.
This aligns with previous genre-specific experiments, which suggest explicitly building an alternate version of embedding similarity rankers that can isolate negative sentiment texts to avoid this issue in decrease tasks \cite{penalozaTEARSTextualRepresentations2024}.

\label{sec:experiment_recommender_method_profile}
We next examine the profile generation component, comparing the steerability of paragraph-length and sentence-length profiles in Table \ref{tab:compare_profiler}.
Both profile types were successfully steered, neither profile format is unilaterally better, and the magnitude of differences is moderate, which suggests that the steerability of a natural language recommendation system tolerates some variation in the original profile text.

\subsubsection{What recommender performs the best?}
\label{sec:experiment_overall_success}

Based on these results, we set a default natural language recommendation pipeline that generates paragraph-length profiles and ranks items by prompting a large language model to predict ratings for each item and sorts items by descending predicted rating.

For all users tested, we found that this default pipeline succeeded at steering the prevalence of each tag upwards or downwards as appropriate, with an average increase $\Delta {AUC}_t$ of 0.0476 and decrease $\Delta {AUC}_t$ of -0.0266 over 75 trigger warning tags and 19 genre tags.
Based on a one-tail paired t-test ($n=940$)\footnote{While we use the same 10 users, we treat each tag and task combination as independent because they differ in steering intervention.}, this was statistically significant for both increase ($p=4.9 \cdot 10^{-73}$) and decrease ($p=3.9 \cdot 10^{-35}$) tasks.
For the increase task, 89 out of 94 total tags were successfully steered (with positive $\Delta {AUC}_t$). For decrease, 76 out of 94 tags were successfully steered (with negative $\Delta {AUC}_t$).
Overall, while this is a small impact, it does show that steering is generally successful across a broad range of tags. In the following we analyze in more detail when and how steering is successful, and why steering fails in some settings.

\begin{table}[]
    \centering
    \begin{tabular}{l|c c}
         & Incr. & Decr. \\
         Steering intervention & $\Delta {AUC}_t$ ($\uparrow$) & $\Delta {AUC}_t$ ($\downarrow$) \\
         \midrule[0.15ex]
         % (Genre) & & \\
         % \textit{Template append} & 0.1095 (*) & \textbf{-0.0654} (*) \\
         % LLM append & 0.0929 (*) & -0.0442 (*) \\
         % LLM rewrite & \textbf{0.1204} (*) & -0.0569 (*) \\
         % \midrule[0.15ex]
         % (Trigger-all) & & \\
         % \textit{Template append} & \textbf{0.0297} (*) & \textbf{-0.0169} (*) \\
         % LLM append & 0.0245 (*) & -0.0123 (*) \\
         % LLM rewrite & 0.0286 (*) & \textbf{-0.0338} (*) \\
         % \midrule[0.15ex]
         % (Trigger) & & \\
         % \textit{Template append} & 0.0320 (*) & \textbf{-0.0168} (*) \\
         % LLM append & 0.0288 (*) & -0.0136 (*) \\
         % LLM rewrite & \textbf{0.0433} (*) & \textbf{-0.0331} (*) \\
         % \midrule[0.15ex]
         % (All-all) & & \\
         % \textit{Template append} & 0.0400 (*) & -0.0231 (*) \\
         % LLM append & 0.0333 (*) & -0.0164 (*) \\
         % LLM rewrite & \textbf{0.0405} (*) & \textbf{-0.0368} (*) \\
         % \midrule[0.15ex]
         % (All) & & \\
         Template append & 0.0476 (*) & -0.0266 (*) \\
         LLM append & 0.0417 (*) & -0.0198 (*) \\
         LLM rewrite & \textbf{0.0589} (*) & \textbf{-0.0379} (*) \\
    \end{tabular}
    \caption{Steering intervention method comparison results. The best-performing method for each task is bolded. Asterisks denote statistically significant steering success based on one-tail paired t-test ($\alpha=0.05$). LLM-rewritten profiles produce the most effective steering.}
    \label{tab:compare_steering}
    \vspace{-2em}
\end{table}

\subsection{How does steerability vary across different intervention methods?}
\label{sec:experiment_steering_method}

Ideally, a natural language recommender should tolerate a variety of possible text edits from the user.
Thus, we investigate how steerability varies across a range of steering interventions.

We compare steering intervention methods (Table \ref{tab:compare_steering}) and find that all of them show some level of success.
This is very encouraging, as it suggests that the steerability of a natural language recommender tolerates variations in how a user revises their profile, in addition to the variations in generated profile format that we found earlier.
Prompting a language model to fully revise the profile may be the most successful because it minimizes the chance of any conflicting information in the original profile remaining after the steering action.
However, fully revising a profile is also the intervention that would require the most effort from an end user, because it requires fully re-reading and editing the text as a whole, instead of simply appending a single instruction.

\subsubsection{How does degree of emphasis in a steering intervention impact steering effectiveness?}
\label{sec:experiment_steering_method_emphasis}

\begin{table}[]
    \centering
    \begin{tabular}{l|c c}
         & Incr. & Decr. \\
         Emphasis level & $\Delta {AUC}_t$ ($\uparrow$) & $\Delta {AUC}_t$ ($\downarrow$) \\
         % \midrule[0.15ex]
         % (Genre) & & \\
         % Stronger & \textbf{0.0998} (*) & \textbf{-0.0923} (*) \\
         % Weaker & 0.0873 (*) & -0.0588 (*) \\
         % \midrule[0.15ex]
         % (Trigger-all) & & \\
         % Stronger & X & X \\
         % Weaker & X & X \\
         % \midrule[0.15ex]
         % (Trigger) & & \\
         % Stronger & \textbf{0.0460} (*) & \textbf{-0.0179} (*) \\
         % Weaker & 0.0369 (*) & -0.0131 (*) \\
         % \midrule[0.15ex]
         % (All-all) & & \\
         % Stronger & X & X \\
         % Weaker & X & X \\
         \midrule[0.15ex]
         % (All) & & \\
         Stronger & \textbf{0.0569} (*) & \textbf{-0.0330} (*) \\
         Weaker & 0.0471 (*) & -0.0224 (*) \\
    \end{tabular}
    \caption{Steering emphasis level comparison results. The best-performing method for each task is bolded. Asterisks denote statistically significant steering success based on one-tail paired t-test ($\alpha=0.05$). As expected, stronger language results in larger changes in the steering responses.}
    \label{tab:compare_emphasis}
    \vspace{-2em}
\end{table}

It is possible that an end user would want to adjust "how much" steering is done -- instead of strictly removing all items that match a condition, they may simply want to see fewer of those items.
We tested whether more strictly phrased steering interventions produced more visibly steered rankings, by comparing a pair of "stronger" emphasis steering instructions against "weaker" emphasis instructions.
Results are shown in Table \ref{tab:compare_emphasis}.
Exact instruction phrasing is presented in Appendix \ref{app:prompts}.
% \begin{itemize}
%     \item Stronger Genre: "Please show the user *only* Romance movies." / "Please show the user *no* Romance movies."
%     \item Weaker Genre: "Please show the user *more* Romance movies." / "Please show the user *less* Romance movies."
%     \item Stronger Trigger: "The user wants to see *only* movies where the dog dies." / "The user wants to see *no* movies where the dog dies."
%     \item Weaker Trigger: "The user wants to see *more* movies where the dog dies." / "The user wants to see *less* movies where the dog dies."
% \end{itemize}
We find that a strongly, strictly worded intervention did produce a stronger steering effect.
This was more visible for the decrease task than the increase task.

\subsection{What topics are easier or harder to steer for?}
\label{sec:experiment_per_topic}

\begin{table}[]
    \centering
    \begin{tabular}{l|c c}
        & Incr. & Decr. \\
        Tag source (\# of tags)& $\Delta {AUC}_t$ ($\uparrow$) & $\Delta {AUC}_t$ ($\downarrow$) \\
        \midrule[0.15ex]
        Genre (19) &  \textbf{0.1095} &  \textbf{-0.0654} \\
        Trigger (75) & 0.0320 & -0.0168 \\
        Trigger (128) & 0.0297 & -0.0169 \\
    \end{tabular}
    \caption{Comparison of steerability between tag types. Note that Genre tags produce the largest response for both increase and decrease steering actions.}
    \label{tab:compare_tag_group}
    \vspace{-2em}
\end{table}

We compare how steerability varies when steering well-known genres versus more nuanced trigger warnings in Table \ref{tab:compare_tag_group}.
Overall, we find that steering was more successful on average for genres than for trigger warnings.

% \todo{vis comparison of movielens, trigger tag steerability results?}

\begin{table}[]
    \centering
    \setlength{\tabcolsep}{3pt}
    \begin{tabular}{>{\footnotesize}l|>{\footnotesize}c|>{\footnotesize}c}
        & {\footnotesize Genre tags ($\Delta {AUC}_t$)} & {\footnotesize Trigger tags ($\Delta {AUC}_t$)} \\
        \midrule[0.15ex]
        Biggest & War (0.1452), & The fourth wall is broken (0.1623), \\
        increase & Mystery (0.1262), & Someone says 'I'll kill myself' (0.1594), \\
        & Sci-Fi (0.1251) & There are babies or unborn children (0.1448) \\
        \midrule[0.15ex]
        Biggest & Documentary (-0.1783) & The fourth wall is broken (-0.1912) \\
        decrease & Animation (-0.1264) & Someone has an eating disorder (-0.1783) \\
        & Fantasy (-0.0836) & Someone says 'I'll kill myself' (-0.1224) \\
        \midrule[0.15ex]
        Smallest & Drama (0.0326) & There is cannibalism (0.0005) \\
        increase & IMAX (0.0393) & Someone miscarries (0.0006) \\
        & Children (0.0490) & A head gets squashed (0.0017) \\
        % \midrule[0.15ex] % version that includes failures to steer
        % Worst incr & Drama (0.0326) & Someone becomes unconscious (-0.0304) \\
        % & IMAX (0.0393) & There are mannequins (-0.0100) \\
        % & Children (0.0490) & There is a large age gap (-0.0037) \\
        \midrule[0.15ex]
        Smallest & Romance (-0.0060) & Someone becomes unconscious (-0.0011) \\
        decrease & War (-0.0230) & The dog dies (-0.0011) \\
        & Western (-0.0240) & Someone drowns (-0.0014) \\
        % \midrule[0.15ex] % version that includes failures to steer
        % Worst decr & IMAX (0.0031) & There is a hanging (0.0368) \\
        % & Romance (-0.0060) & There is a dead animal (0.0259) \\
        % & War (-0.0230) & A child ia abandoned by a parent (0.0163) \\
    \end{tabular}
    
    \caption{
    Average $\Delta$AUC per tag for the top 3 most and least successfully steered genre and trigger warning tags.
    Trigger tags that were frequently rejected by the LLM or failed to steer are excluded.
    }
    \label{tab:sample_tags}
    \vspace{-2em}
\end{table}

\label{sec:experiment_per_topic_variation}
In addition, we found some variation in how steerable each tag was within the topic groups.
% There is some variation in how frequently tags are ranked highly before any steering intervention at all, which may affect how much they can be steered.
We visualize a sample of the most and least successfully steered genre and trigger warning tags in Table \ref{tab:sample_tags}.

\subsubsection{LLM Safety}
\label{sec:experiment_per_topic_safety}

Notably, when prompting a language model to edit an existing profile, the language model sometimes outputs a refusal message or misinterprets the tag name.
This happens especially frequently when prompting the model to increase tag prevalence for any trigger warning related to prejudice, graphic imagery, or certain mental health topics, and is likely caused by model safety training in pretrained models.

To mitigate this effect, one author annotated and filtered out 53 tags where over one quarter of LLM responses for the increase or decrease task were grossly misinterpreted or contained some refusal text.
Aggregate results based on the remaining 75 tags are labeled as "Trigger (75)" in Table \ref{tab:compare_tag_group}, while the full list of trigger warnings including frequently-rejected tags is labeled as "Trigger (128)".
% Note that the overall steering effectiveness on the filtered "Trigger" tags is better for the increase task than the decrease task.
% This is likely because many removed tags were successfully steered for the decrease task, but frequently rejected and had low random failure rates for the increase task.

\begin{table}[]
    \centering
    \begin{tabular}{l |c c}
         & Incr. & Decr. \\
         Condition & $\Delta {AUC}_t$ ($\uparrow$) & $\Delta {AUC}_t$ ($\downarrow$) \\
         \midrule[0.15ex]
         Genre-Oracle & \textbf{0.1660} (*) & \textbf{-0.1504} (*) \\
         Genre & 0.1095 (*) & -0.0654 (*) \\
         % \midrule[0.15ex]
         % Trigger-all-Oracle & \textbf{0.1162} (*) & \textbf{-0.1727} (*) \\
         % Trigger-all & 0.0297 (*) & -0.0169 (*) \\
         \midrule[0.15ex]
         Trigger-Oracle & \textbf{0.1211} (*) & \textbf{-0.1721} (*) \\
         Trigger & 0.0320 (*) & -0.0168 (*) \\
         % \midrule[0.15ex]
         % All-all-Oracle & \textbf{0.1226} (*) & \textbf{-0.1698} (*) \\
         % All-all & 0.0400 (*) & -0.0231 (*) \\
         \midrule[0.15ex]
         All-Oracle & \textbf{0.1302} (*) & \textbf{-0.1677} (*) \\
         All & 0.0476 (*) & -0.0266 (*) \\
    \end{tabular}
    \caption{Comparison of how adding oracle information to item descriptions affected steering success. The best-performing method for each task and tag group is bolded. Asterisks denote statistically significant steering success based on one-tail paired t-test ($\alpha = 0.05$). Adding information about the steered tag to item metadata text significantly improves steering effectiveness. The effect is stronger for trigger warning tags.}
    \label{tab:compare_tag_group_oracle}
    \vspace{-2em}
\end{table}

\subsubsection{Does augmenting a language model with world knowledge improve steerability?}
\label{sec:experiment_per_topic_oracle}

We were curious how much of the steering effectiveness disparity between genre and trigger warning tags was due to the language model lacking world knowledge and being unable to identify whether or not a movie was relevant.
% based on its title and description.
To test this, we compared our default pipeline against an oracle ranker, where the item metadata used for ranking includes information about whether the tag being steered pertains to each individual item.
Results are shown in Table \ref{tab:compare_tag_group_oracle}.

We found that this oracle ranker performed significantly and substantially better than the regular rankings for both types of tags.
Furthermore, it provides greater improvements for the trigger warning tags than it does for the genre tags.
This suggests that some steering failures or larger disparities in how easy topics are to steer are caused by language model knowledge gaps - that is, the recommender is unable to steer because it simply lacks knowledge or reasoning for whether a given topic is relevant to the item.
Based on this, it might be possible to improve a natural language recommender by extracting or reasoning about preference-related metadata and using that to improve ranking. We discuss this further in section \ref{sec:discussion}.

% \notejz{A more detailed table of tag steering success, rejection, and oracle improvement rates is available in Appendix \ref{app:all_tags}.}

\subsection{How does steering affect accuracy?}
\label{sec:experiment_accuracy}

\begin{table}[]
    \centering
    \begin{tabular}{l l|c c}
         Task & Type & Original ${Rank}_{set}$ & $\Delta {Rank}_{set}$ \\
         \midrule[0.15ex]
         Increase & Rel. & 21.09 & 3.38 \\
         Increase & Irrel. & 19.37 & -0.98 \\
         Decrease & Rel. & 20.47 & -0.58 \\
         Decrease & Irrel. & 19.15 & 0.05 \\
    \end{tabular}
    \caption{Comparison of how steering impacted ranking of the true next movie compared to the other 49 relevant or irrelevant movie subset. Lower ${Rank}_{set}$ means better ranking visibility.}
    \label{tab:accuracy_vs_steering}
    \vspace{-2em}
\end{table}

Finally, we evaluate the trade-offs exist between recommendation accuracy and steering.

Traditionally, recommendation accuracy is judged based on whether a recommender can correctly predict the next item that a user will interact with.
An accurate natural language profile is one that can suggest the correct item.
Any steering intervention introduces noise to this profile.
For example, in an increase task, this might harm accuracy by encouraging new items that are relevant to the tag to become ranked more highly than the original correct item.
In a decrease task, this might harm accuracy if the correct item is also relevant to the tag and therefore ranked more lowly.
Ideally, even if general accuracy decreases, the ranking of the correct item within the subset of items that are similarly relevant or irrelevant to the targeted tag should still stay unchanged: a steering action can impact the subset as a whole, but should not forget critical information about a user's preferences from before.

We define a relative subset ranking metric ${Rank}_{set}$ that captures this notion.
If the correct item is irrelevant to the tag, ${Rank}_{set}$ denotes its ranking within the 50 total items that are irrelevant to the tag, where lower ${Rank}_{set}$ means it has a higher position in the ranking.
If the correct item is relevant to the tag, ${Rank}_{set}$ denotes its ranking within the 50 total items that are relevant to the tag.

In Table \ref{tab:accuracy_vs_steering}, we sample 100 steering scenarios for each possible steering action and next-item tag relevance condition and show the average $\Delta {Rank}_{set}$ for the default recommender pipeline.
We find that the impact on the target item's ranking relative to a set of other items that should be equally impacted is small, indicating that prior preference information is retained as desired.
We discuss additional approaches to maintaining recommendation accuracy in section \ref{sec:discussion}.

\section{Discussion}
\label{sec:discussion}

We start by highlighting several practical design suggestions based on our experiments.
First, \textit{well-known topics (such as genre) are easier to steer than more niche topics}. This is likely due to lack of language model world knowledge or reasoning. This could potentially be ameliorated by augmenting the item metadata used for ranking.
We did observe that when we added explicit information about steered topics to the metadata for each item in the oracle condition, steerability significantly improved.
This suggests that it might be possible to improve a system by explicitly extracting metadata relevant to any steering actions for each item, or otherwise improving the metadata or prompts that are used in the ranking components.

Second, although an embedding-based ranker is faster, \textit{pretrained embedding models are difficult to meaningfully steer for decrease tasks}.
While there have been methods proposed \cite{penalozaTEARSTextualRepresentations2024} to sidestep the embedding model challenges, these require explicitly identifying a negative sentiment and extracting a steered topic from the underlying profile, which may be difficult depending on interface design and user steering behaviors.

Third, \textit{natural language profiles are fairly tolerant of different intervention methods} (barring steering requests that conflict with language model safety guardrails).
A full profile rewrite is generally more effective than simpler sentence-appending interventions. Steering is also sensitive to the degree of emphasis used in profile revision.

Finally, while there is a tradeoff between steering effectiveness and overall accuracy as expected, \textit{recommenders largely preserve accuracy when evaluated orthogonally to the steered topics}.
Depending on how recommendations are generated and presented to an end user, this challenge can also be avoided -- for example, a system could explicitly interweave accuracy-dedicated recommendations with steered recommendations.

% \jznote{Do we have any additional mild observations in the types of profiles and interventions that were generated?}

% \jznote{What limitations are there in our evaluation tool? How can this tool be extended to other domains or recommendation setups?}

However, it is worth noting that while we evaluate a broader set of topics that people might steer on, the number of possible steering request types or topics is vast and any dataset of benchmark steering tags is limited.
For example, a user could request "more intellectually stimulating recommendations", "less of what's currently popular", or "do not encourage me to watch more than two movies per week", which are outside our evaluation scope.

Furthermore, we do not finetune any machine learning models that might contribute towards improved accuracy or steering effectiveness.
It is possible that finetuning any model component involved in profile creation, steering intervention, or ranking would improve steering or address some of the challenges we observed in our experiments, such as topic rejection or embedding model failure.
However, we consider this model improvement beyond the scope of this work.
In the future, the \sysname\ framework can be adapted to evaluate finetuned models with additional task datasets or steering interventions as appropriate.

Finally, there are multiple larger human-recommender interactions that we do not account for and intentionally exclude from this evaluation framework due to design limitations.
For instance, we do not account for how changes in user behavior over time may impact steerability, nor do we explore the possibilities of measuring the impact or persistance of any given steering action over time as a user continues to use a recommender.

In addition, it takes effort for an end user to read and edit a natural language profile.
In our experiments, we do not examine any trade-off between how easy such steering actions may be for a human to do and how effective the steering action is.
The user interface design for a steerable natural language profile recommender might also impact overall effectiveness.
Future work could investigate factors such as cognitive load and profile transparency as they relate to the design of steerable recommender system user interfaces.

\begin{acks}
This research was supported by collaborative NSF Award IIS-2312865 / IIS-2312866, NSF Award OAC-2311521, and in part by the Graduate Fellowships for STEM Diversity (GFSD).
All content represents the opinion of the authors, which is not necessarily shared or endorsed by their respective employers and/or sponsors.
\end{acks}

\bibliographystyle{ACM-Reference-Format}
\bibliography{references}

\appendix

\section{Replication of previous steerability evaluations}
\setcounter{table}{0}
\renewcommand{\thetable}{A\arabic{table}}
\label{app:replication}

\begin{table}[h]
    \centering
    \setlength{\tabcolsep}{6pt}
    \renewcommand{\arraystretch}{1.1}
    \footnotesize
    \begin{tabular}{l|l|p{0.55\linewidth}}
        Method & Tags & $\Delta$ Coverage@10 \\
        \midrule[0.15ex]
        UCL 
        & 2 genres 
        & Increase: +0.20 (Comedy)
        \newline
         Increase: +0.35 (Horror) \\
        \midrule[0.15ex]
        Ours 
        & 19 genre 
        & Increase: +0.2021;\newline
          Decrease: $-0.0810$ \\
    \end{tabular}
    
    \caption{
    Contextual replication of UCL-style tag steering under Coverage@10. While setups differ, gains under tag/genre increase are directionally consistent.
    }
    \label{tab:ucl_replication}
    \vspace{-2em}
\end{table}

\begin{table}[h]
    \centering
    \setlength{\tabcolsep}{6pt}
    \renewcommand{\arraystretch}{1.1}
    \footnotesize
    \begin{tabular}{l|l|l|p{0.48\linewidth}}
        Method & Steering & Tags & $\Delta$ NDCG \\
        \midrule[0.15ex]
        TEARS
        & Increase
        & 3 genres
        & NDCG@20: +0.245 to +0.280\newline
          (full profile rewrite) \\
        \midrule[0.15ex]
        TEARS
        & Decrease
        & 3 genres
        & NDCG@20: -0.245 to -0.280\newline
          (full profile rewrite)\newline
          (custom decrease implementation) \\
        \midrule[0.15ex]
        Ours
        & Increase
        & 19 genres
        & NDCG@10: +0.2124 \newline
          NDCG@20: +0.1936 \\
        \midrule[0.15ex]
        Ours
        & Decrease
        & 19 genres
        & NDCG@10: $-0.0755$ \newline
          NDCG@20: $-0.0594$ \\
    \end{tabular}
    
    \caption{
    Contextual comparison with TEARS. Larger-scope profile rewrites yield larger gains, while our smaller-scope genre steering produces consistent but more conservative improvements.
    }
    \label{tab:tears_replication}
    \vspace{-2em}
\end{table}

Previous work in natural language profile recommender steerability evaluation has used varying steerability evaluation metrics and ranking sizes.
While we do not use the exact same metrics and ranking sizes, we do generally replicate their findings and discuss that here.

UCL \cite{ramosTransparentScrutableRecommendations2024} evaluated a version of tag increase on the movie (Amazon-MT) and travel (TripAdvisor) domains with 2 tags each, measuring Coverage@10 for a ranking pool containing 100 relevant items and 100 irrelevant items. We compare their findings with our measurements of Coverage@10 in Table \ref{tab:ucl_replication}.

TEARS \cite{penalozaTEARSTextualRepresentations2024} evaluated a version of tag increase on the movie (MovieLens-1M and Netflix) and books (Goodbooks) domains for 3 genre tags each, measuring NDCG@20 for a ranking pool containing all rankable items for each domain. We compare their findings with our measurements of NDCG in Table \ref{tab:tears_replication}.

\section{Prompts and Example Generations}
\label{app:prompts}

% Tables \ref{appx:tab:template_append}, \ref{appx:tab:template_blurb_oracle}, \ref{appx:tab:prompt_profile_sentence}, \ref{appx:tab:prompt_profile_paragraph}, \ref{appx:tab:prompt_revise_append}, \ref{appx:tab:prompt_revise_rewrite}, \ref{appx:tab:prompt_score}.

\begin{table}[h]\centering
\raggedright{\textbf{Append Templates (+ Varying Emphasis)}}
\resizebox{\linewidth}{!}{
\begin{tabular}{p{1.3\linewidth}} 
\midrule[0.3ex]
Please show the user movies that satisfy: \textit{Mystery}.\newline
The user *does not want* to see \textit{Mystery}.\newline
\newline
The user *wants* to see movies where \textit{there are ghosts}.\newline
The user *does not want* movies where \textit{there are ghosts}.\newline
\newline
Please show the user movies *only* \textit{Mystery} movies.\newline
Please show the user movies *more* \textit{Mystery} movies.\newline
Please show the user movies *less* \textit{Mystery} movies.\newline
Please show the user movies *no* \textit{Mystery} movies.\newline
\newline
The user wants to see *only* movies where \textit{there are ghosts}.\newline
The user wants to see *more* movies where \textit{there are ghosts}.\newline
The user wants to see *less* movies where \textit{there are ghosts}.\newline
The user wants to see *no* movies where \textit{there are ghosts}.\\
\midrule[0.3ex]
\end{tabular}}
\label{appx:tab:template_append}
\end{table}

\begin{table}[h]\centering
\raggedright{\textbf{Item Metadata Blurb - Oracle Condition}}
\resizebox{\linewidth}{!}{
\begin{tabular}{p{1.3\linewidth}} 
\midrule[0.3ex]
Movie Title: The Stranger\newline
Movie Description: An investigator from the War Crimes Commission  travels to Connecticut to find an infamous $\cdots$ Supreme Court Justice’s daughter.\\
% Movie Description: An investigator from the War Crimes Commission travels to Connecticut to find an infamous Nazi, who may be hiding out in a small town in the guise of a distinguished professor engaged to the Supreme Court Justice’s daughter.\\
\midrule[0.3ex]
\end{tabular}}
\label{appx:tab:template_blurb_oracle}
\end{table}

\begin{table}[h]\centering
\raggedright{\textbf{Profile Generation (Paragraph)}}
\resizebox{\linewidth}{!}{
\begin{tabular}{p{1.3\linewidth}} 
\midrule[0.3ex]
Given the user's previously watched and rated movies, write in a single paragraph (5-6 sentences) summarizing their taste. Be definitive (no hedging). Do NOT mention tags, genres, metadata, reviews, or any lack thereof. Infer tone, pacing, themes, and style.\newline
\newline
User watch history:\newline
\newline
Movie Title: Jumanji\newline
Movie Description: When siblings Judy and Peter discover an enchanted board game $\cdots$ rhinoceroses, evil monkeys and other terrifying creatures.\newline
% Movie Description: When siblings Judy and Peter discover an enchanted board game that opens the door to a magical world, they unwittingly invite Alan -- an adult who's been trapped inside the game for 26 years -- into their living room. Alan's only hope for freedom is to finish the game, which proves risky as all three find themselves running from giant rhinoceroses, evil monkeys and other terrifying creatures.\newline
User Rating: 4.0\newline
\newline
$\cdots$\\
\midrule[0.3ex]
This user has a strong affinity for timeless, character-driven stories that often blend elements of drama, adventure, and fantasy. They appreciate complex, emotionally resonant narratives that explore the human condition, frequently set against the backdrop of historical or social upheaval. These stories often feature memorable, flawed characters who must navigate their circumstances with wit, courage, and heart. They enjoy films that balance action, suspense, and humor, and frequently feature elements of romance and redemption. Their tastes lean towards the classic and the epic, with a fondness for period dramas, historical epics, and coming-of-age stories.\\
\midrule[0.3ex]
\end{tabular}}
\label{appx:tab:prompt_profile_paragraph}
\end{table}

\begin{table}[h]\centering
\raggedright{\textbf{Profile Generation (sentence)}}
\resizebox{\linewidth}{!}{
\begin{tabular}{p{1.3\linewidth}} 
\midrule[0.3ex]
Given the user's previously watched and rated movies, write exactly ONE concise English sentence (20-35 words) summarizing their taste. Be definitive (no hedging). Do NOT mention tags, genres, metadata, reviews, or any lack thereof. Infer tone, pacing, themes, and style.\newline
\newline
$\cdots$\\
\midrule[0.3ex]
This user has a strong affinity for timeless, character-driven stories that often involve themes of redemption, love, and the human condition, frequently set against a backdrop of adventure, drama, and wit.\\
\midrule[0.3ex]
\end{tabular}}
\label{appx:tab:prompt_profile_sentence}
\end{table}

\begin{table}[h]\centering
\raggedright{\textbf{Append LLM-written Text (Trigger decrease example)}}
\resizebox{\linewidth}{!}{
\begin{tabular}{p{1.3\linewidth}} 
\midrule[0.3ex]
You are helping a user update how they describe their movie preferences by rephrasing their changed preferences into a single sentence. This sentence will be appended to their original description.\newline
Guidelines:\newline
- Write exactly *one* sentence. The sentence must start with "The user".\newline
- Do not include any "REQUEST" in the response. Only give the rephrased sentence.\newline
- Use natural language.\newline
\newline
For example:\newline
REQUEST: The user *does not want* to see movies where [the dog dies].\newline
The user *does not want* to see any movies where a dog dies.\newline
\newline
REQUEST: The user *does not want* to see movies where \textit{there are ghosts}.\\
\midrule[0.3ex]
The user does not want to see movies where there are ghosts.\\
\midrule[0.3ex]
\end{tabular}}
\label{appx:tab:prompt_revise_append}
\end{table}

% \newpage

\begin{table}[h]\centering
\raggedright{\textbf{Rewrite with LLM (Trigger decrease example)}}
\resizebox{\linewidth}{!}{
\begin{tabular}{p{1.3\linewidth}} 
\midrule[0.3ex]
Modify the user profile to show that the user *does not want* to see movies where \textit{there are ghosts}.\newline
Keep all the profile as similar as possible for all other preferences.\newline
\newline
ORIGINAL PROFILE: \textit{The user has a strong affinity for timeless, $\cdots$ historical epics, and coming-of-age stories.}\newline
UPDATED PROFILE:\\
\midrule[0.3ex]
This user has a strong affinity for timeless, character-driven stories that often blend elements of drama, adventure, and fantasy. They appreciate complex, emotionally resonant narratives that explore the human condition, frequently set against the backdrop of historical or social upheaval. These stories often feature memorable, flawed characters who must navigate their circumstances with wit, courage, and heart. They enjoy films that balance action, suspense, and humor, and frequently feature elements of romance and redemption. However, they have a preference to avoid movies that involve supernatural elements, such as ghosts, and instead lean towards classic and epic stories like period dramas, historical epics, and coming-of-age tales.\\
\midrule[0.3ex]
\end{tabular}}
\label{appx:tab:prompt_revise_rewrite}
\end{table}

\begin{table}[h]\centering
\raggedright{\textbf{Score prediction with LLM}}
\resizebox{\linewidth}{!}{
\begin{tabular}{p{1.3\linewidth}} 
\midrule[0.3ex]
Task: Predict the user's rating for a movie on a scale from 0.0 to 5.0\newline
Guidelines:\newline
- 0.0: Completely irrelevant.\newline
- 5.0: Perfect match for ALL user interests.\newline
- Use decimal precision as necessary (e.g. 1.4, 3.7, 4.2) to capture partial matches.\newline
- Be strict. Do not give high scores for partial matches.\newline
Output format: A single float number only.\newline
\newline
USER PROFILE:\newline
\textit{The user has a strong affinity for timeless, $\cdots$ historical epics, and coming-of-age stories.}\newline
CANDIDATE MOVIE:\newline
Movie Title: The Stranger\newline
Movie Description: An investigator from the War Crimes Commission  travels to Connecticut to find an infamous $\cdots$ Supreme Court Justice’s daughter.\newline
% Movie Description: An investigator from the War Crimes Commission travels to Connecticut to find an infamous Nazi, who may be hiding out in a small town in the guise of a distinguished professor engaged to the Supreme Court Justice’s daughter.\newline
SCORE:\\
\midrule[0.3ex]
3.4\newline
\newline
The movie "The Stranger" appears to match the user's profile in several key aspects. $\cdots$ However, the movie's focus on a thriller plot and the absence of a strong emphasis on romance, redemption, and epic scope might prevent it from being a perfect match.\\
\midrule[0.3ex]
\end{tabular}}
\label{appx:tab:prompt_score}
\end{table}

% \section{Steerability of all genre and trigger warning tags}
% \label{app:all_tags}

% \todo{big table with increase, decrease, llm responsiveness reliability annotation for each tag??????}

\end{document}